\documentclass[12pt,oneside,reqno,american]{amsart}
\usepackage{biolinum}
\usepackage[T1]{fontenc}
\usepackage[latin9]{inputenc}
\usepackage{color}
\definecolor{lyxboxbgcolor}{rgb}{1, 0, 0}
\definecolor{lyxnotefontcolor}{rgb}{0.75, 0.75, 0.75}
\usepackage{array}
\usepackage{verbatim}
\usepackage{amsbsy}
\usepackage{amstext}
\usepackage{amsthm}
\usepackage{amssymb}
\usepackage{xargs}[2008/03/08]

\makeatletter

\providecommand{\tabularnewline}{\\}

\numberwithin{equation}{section}
\numberwithin{figure}{section}
\numberwithin{table}{section}

\usepackage{mathtools}
\usepackage{amsthm}
\usepackage{pdfsync} 
\usepackage{hyperref}
\usepackage[all]{xy}

\usepackage[T1]{fontenc}
\usepackage{Alegreya} 
\usepackage[stix2,scaled=0.96]{newtxmath}
\usepackage[cal=boondoxo,bb=boondox,frak=boondox,scaled=1.10]{mathalfa}
\usepackage[scaled=.9]{rsfso}
\usepackage{xcolor} 
\definecolor{brown(traditional)}{rgb}{0.59, 0.29, 0.0}
\definecolor{blue(ryb)}{rgb}{0.01, 0.28, 1.0}
\definecolor{red}{rgb}{1.0, 0.0, 0.0}
\definecolor{magenta}{rgb}{1.0, 0.0, 1.0}
\definecolor{mahogany}{rgb}{0.75, 0.25, 0.0}
\definecolor{lavenderpurple}{rgb}{0.59, 0.48, 0.71}
\definecolor{olive}{rgb}{0.5, 0.5, 0.0}
\definecolor{brickred}{rgb}{0.8, 0.25, 0.33}
\definecolor{antiquefuchsia}{rgb}{0.57, 0.36, 0.51}
\definecolor{bole}{rgb}{0.47, 0.27, 0.23}
\definecolor{darkolivegreen}{rgb}{0.33, 0.42, 0.18}
\definecolor{deepjunglegreen}{rgb}{0.0, 0.29, 0.29}
\definecolor{brickred}{rgb}{0.8, 0.25, 0.33}
\definecolor{deepjunglegreen}{rgb}{0.0, 0.29, 0.29}
\definecolor{darkpastelgreen}{rgb}{0.01, 0.75, 0.24}
\definecolor{green(pigment)}{rgb}{0.0, 0.65, 0.31}
\definecolor{junglegreen}{rgb}{0.16, 0.67, 0.53}
\definecolor{officegreen}{rgb}{0.0, 0.5, 0.0}
\definecolor{seagreen}{rgb}{0.18, 0.55, 0.34}
\definecolor{teal}{rgb}{0.0, 0.5, 0.5}
\definecolor{brightgreen}{rgb}{0.4, 1.0, 0.0}
\definecolor{electricgreen}{rgb}{0.0, 1.0, 0.0}
\definecolor{malachite}{rgb}{0.04, 0.85, 0.32}

\usepackage{accents}
\usepackage{colortbl}
\usepackage{verbatim}
\usepackage{float}
\usepackage{booktabs}
\usepackage{tikz}
\usetikzlibrary{arrows.meta,positioning}
\tikzset{
  negchg/.style={circle,draw,fill=cyan!8,inner sep=2.2pt},
  poschg/.style={circle,draw,fill=red!25,inner sep=2.2pt},
  vec/.style={-{Stealth[length=2.2mm]},thick},
  frc/.style={-{Stealth[length=2.2mm]},thick,red!70!black},
  vel/.style={-{Stealth[length=2.2mm]},thick,blue!80!black}
}

\makeatother

\providecommand\theoremname{Theorem}
\theoremstyle{plain}
\newtheorem{thm}{\protect\theoremname}[section]
\providecommand\propositionname{Proposition}
\newtheorem{prop}[thm]{\protect\propositionname}
\providecommand\definitionname{Definition}
\theoremstyle{definition}
\newtheorem{defn}[thm]{\protect\definitionname}
\providecommand\remarkname{Remark}
\theoremstyle{remark}
\newtheorem{rem}[thm]{\protect\remarkname}
\providecommand\examplename{Example}
\theoremstyle{definition}
\newtheorem{example}[thm]{\protect\examplename}
\usepackage{babel}
\begin{document}

\global\long\def\ga{\alpha}%
\global\long\def\gb{\beta}%
\global\long\def\ggm{\gamma}%
\global\long\def\go{\omega}%
\global\long\def\gs{\sigma}%
\global\long\def\gd{\delta}%
\global\long\def\gD{\Delta}%
\global\long\def\vph{\phi}%
\global\long\def\gf{\phi}%
\global\long\def\gk{\kappa}%
\global\long\def\gl{\lambda}%
\global\long\def\gz{\zeta}%
\global\long\def\gh{\eta}%
\global\long\def\gy{\upsilon}%
\global\long\def\gth{\theta}%
\global\long\def\gO{\Omega}%
\global\long\def\gG{\Gamma}%

\global\long\def\eps{\varepsilon}%
\global\long\def\epss#1#2{\varepsilon^{#1}_{#2}}%
\global\long\def\ep#1{\eps_{#1}}%

\global\long\def\wh#1{\widehat{#1}}%
\global\long\def\hi{\hat{\imath}}%
\global\long\def\hj{\hat{\jmath}}%
\global\long\def\hk{\hat{k}}%
\global\long\def\ol#1{\overline{#1}}%
\global\long\def\ul#1{\underline{#1}}%

\global\long\def\spec#1{\textsf{#1}}%

\global\long\def\v#1{\boldsymbol{#1}}%

\global\long\def\ui{\wh{\boldsymbol{\imath}}}%
\global\long\def\uj{\wh{\boldsymbol{\jmath}}}%
\global\long\def\uk{\widehat{\boldsymbol{k}}}%

\global\long\def\uI{\widehat{\mathbf{I}}}%
\global\long\def\uJ{\widehat{\mathbf{J}}}%
\global\long\def\uK{\widehat{\mathbf{K}}}%

\global\long\def\mc#1{\mathcal{#1}}%
\global\long\def\bs#1{\boldsymbol{#1}}%
\global\long\def\vect#1{\mathbf{#1}}%
\global\long\def\bi#1{\textbf{\emph{#1}}}%

\global\long\def\uv#1{\widehat{\boldsymbol{#1}}}%
\global\long\def\cross{\times}%

\global\long\def\di{d}%
\global\long\def\dee#1{\mathop{d#1}}%

\global\long\def\ddt{\frac{\dee{}}{\dee t}}%
\global\long\def\dbyd#1{\frac{\dee{}}{\dee{#1}}}%
\global\long\def\dby#1#2{\frac{\partial#1}{\partial#2}}%
\global\long\def\dxdt#1{\frac{\dee{#1}}{\dee t}}%

\global\long\def\vct#1{\bs{#1}}%

\global\long\def\partialby#1#2{\frac{\partial#1}{\partial x^{#2}}}%
\newcommandx\parder[2][usedefault, addprefix=\global, 1=]{\frac{\partial#2}{\partial#1}}%
\global\long\def\supdot{^{\!\bs{\mathord{\cdot}}}}%

\global\long\def\fall{,\quad\text{for all}\quad}%

\global\long\def\reals{\mathbb{R}}%

\global\long\def\rthree{\reals^{3}}%
\global\long\def\rsix{\reals^{6}}%
\global\long\def\rn{\reals^{n}}%
\global\long\def\eucl{\mathbb{E}}%
\global\long\def\euthree{\eucl^{3}}%
\global\long\def\euln{\eucl^{n}}%

\global\long\def\prn{\reals^{n+}}%
\global\long\def\nrn{\reals^{n-}}%
\global\long\def\cprn{\overline{\reals}^{n+}}%
\global\long\def\cnrn{\overline{\reals}^{n-}}%
\global\long\def\rt#1{\reals^{#1}}%
\global\long\def\rtw{\reals^{12}}%

\global\long\def\les{\leqslant}%
\global\long\def\ges{\geqslant}%

\global\long\def\dX{\dee{\bp}}%
\global\long\def\dx{\dee x}%
\global\long\def\D{D}%

\global\long\def\from{\colon}%
\global\long\def\tto{\longrightarrow}%
\global\long\def\lmt{\longmapsto}%
\global\long\def\lhr{\lhook\joinrel\longrightarrow}%
\global\long\def\mto{\mapsto}%

\global\long\def\abs#1{\left|#1\right|}%

\global\long\def\isom{\cong}%

\global\long\def\comp{\circ}%

\global\long\def\cl#1{\overline{#1}}%

\global\long\def\fun{\varphi}%

\global\long\def\interior{\textrm{Int}\,}%
\global\long\def\inter#1{\kern0pt  #1^{\mathrm{o}}}%
\global\long\def\interior{\textrm{Int}\,}%
\global\long\def\inter#1{\kern0pt  #1^{\mathrm{o}}}%
\global\long\def\into{\mathrm{o}}%

\global\long\def\sign{\textrm{sign}\,}%
\global\long\def\sgn#1{(-1)^{#1}}%
\global\long\def\sgnp#1{(-1)^{\abs{#1}}}%

\global\long\def\du#1{#1^{*}}%

\global\long\def\tsum{{\textstyle \sum}}%
\global\long\def\lsum{{\textstyle \sum}}%

\global\long\def\dimension{\textrm{dim}\,}%

\global\long\def\esssup{\textrm{ess}\,\sup}%

\global\long\def\ess{\textrm{{ess}}}%

\global\long\def\kernel{\mathop{\textrm{\textup{Kernel}}}}%

\global\long\def\support{\mathop{\textrm{\textup{supp}}}}%

\global\long\def\image{\mathop{\textrm{\textup{Image}}}}%

\global\long\def\diver{\mathop{\textrm{\textup{div}}}}%

\global\long\def\spanv{\textrm{span}}%

\global\long\def\tr{\mathop{\textrm{\textup{tr}}}}%
\global\long\def\tran{\mathrm{tr}}%

\global\long\def\opt{\mathrm{opt}}%

\global\long\def\incl{\mathcal{I}}%
\global\long\def\iden{\imath}%
\global\long\def\idnt{\textrm{Id}}%
\global\long\def\rest{\rho}%
\global\long\def\extnd{e_{0}}%

\global\long\def\proj{\textrm{pr}}%

\global\long\def\L#1{L\bigl(#1\bigr)}%
\global\long\def\LS#1{L_{S}\bigl(#1\bigr)}%

\global\long\def\ino#1{\int_{#1}}%

\global\long\def\half{\frac{1}{2}}%
\global\long\def\shalf{{\scriptstyle \half}}%
\global\long\def\third{\frac{1}{3}}%

\global\long\def\empt{\varnothing}%

\global\long\def\innp#1#2{\left\langle #1,#2\right\rangle }%

\global\long\def\resto#1{|_{#1}}%
\global\long\def\compat#1#2{\left.#1\right|_{#2}}%

\global\long\def\paren#1{\left(#1\right)}%
\global\long\def\bigp#1{\bigl(#1\bigr)}%
\global\long\def\biggp#1{\biggl(#1\biggr)}%
\global\long\def\Bigp#1{\Bigl(#1\Bigr)}%

\global\long\def\braces#1{\left\{  #1\right\}  }%
\global\long\def\sqbr#1{\left[#1\right]}%
\global\long\def\anglep#1{\left\langle #1\right\rangle }%

\global\long\def\bigabs#1{\bigl|#1\bigr|}%
\global\long\def\dotp#1{#1^{\centerdot}}%
\global\long\def\pdot#1{#1^{\bs{\!\cdot}}}%

\global\long\def\eq{\sim}%
\global\long\def\quot{/\!\!\eq}%
\global\long\def\by{\!/\!}%

\global\long\def\stp{\text{{\small \ensuremath{\bigodot}}}}%
\global\long\def\tp{\text{{\small \ensuremath{\bigotimes}}}}%

\global\long\def\mi#1{#1}%
\global\long\def\mii{I}%
\global\long\def\mie#1#2{#1_{1}\cdots#1_{#2}}%

\global\long\def\smi#1{\boldsymbol{#1}}%
\global\long\def\asmi#1{#1}%
\global\long\def\ordr#1{\left\langle #1\right\rangle }%

\global\long\def\symm#1{\paren{#1}}%
\global\long\def\smtr{\mathcal{S}}%

\global\long\def\perm{p}%
\global\long\def\sperm{\mathcal{P}}%

\global\long\def\oneto{1,\dots,}%

\global\long\def\lisub#1#2#3{#1_{1}#2\dots#2#1_{#3}}%

\global\long\def\lisup#1#2#3{#1^{1}#2\dots#2#1^{#3}}%

\global\long\def\lisubb#1#2#3#4{#1_{#2}#3\dots#3#1_{#4}}%

\global\long\def\lisubbc#1#2#3#4{#1_{#2}#3\cdots#3#1_{#4}}%

\global\long\def\lisubbwout#1#2#3#4#5{#1_{#2}#3\dots#3\widehat{#1}_{#5}#3\dots#3#1_{#4}}%

\global\long\def\lisubc#1#2#3{#1_{1}#2\cdots#2#1_{#3}}%

\global\long\def\lisupc#1#2#3{#1^{1}#2\cdots#2#1^{#3}}%

\global\long\def\lisupp#1#2#3#4{#1^{#2}#3\dots#3#1^{#4}}%

\global\long\def\lisuppc#1#2#3#4{#1^{#2}#3\cdots#3#1^{#4}}%

\global\long\def\lisuppwout#1#2#3#4#5#6{#1^{#2}#3#4#3\wh{#1^{#6}}#3#4#3#1^{#5}}%

\global\long\def\lisubbwout#1#2#3#4#5#6{#1_{#2}#3#4#3\wh{#1}_{#6}#3#4#3#1_{#5}}%

\global\long\def\lisubwout#1#2#3#4{#1_{1}#2\dots#2\widehat{#1}_{#4}#2\dots#2#1_{#3}}%

\global\long\def\lisupwout#1#2#3#4{#1^{1}#2\dots#2\widehat{#1^{#4}}#2\dots#2#1^{#3}}%

\global\long\def\lisubwoutc#1#2#3#4{#1_{1}#2\cdots#2\widehat{#1}_{#4}#2\cdots#2#1_{#3}}%

\global\long\def\twp#1#2#3{\dee{#1}^{#2}\wedge\dee{#1}^{#3}}%

\global\long\def\thp#1#2#3#4{\dee{#1}^{#2}\wedge\dee{#1}^{#3}\wedge\dee{#1}^{#4}}%

\global\long\def\fop#1#2#3#4#5{\dee{#1}^{#2}\wedge\dee{#1}^{#3}\wedge\dee{#1}^{#4}\wedge\dee{#1}^{#5}}%

\global\long\def\idots#1{#1\dots#1}%
\global\long\def\icdots#1{#1\cdots#1}%

\global\long\def\norm#1{\|#1\|}%

\global\long\def\nonh{\heartsuit}%

\global\long\def\nhn#1{\norm{#1}^{\nonh}}%

\global\long\def\bigmid{\,\bigl|\,}%

\global\long\def\trps{^{{\scriptscriptstyle \textsf{T}}}}%

\global\long\def\testfuns{\mathcal{D}}%

\global\long\def\ntil#1{\tilde{#1}{}}%

\global\long\def\pis{y}%
\global\long\def\xo{\pis_{0}}%
\global\long\def\x{x}%

\global\long\def\pib{x}%
\global\long\def\bp{X}%
\global\long\def\ii{i}%
\global\long\def\ia{\alpha}%
\global\long\def\fp{y}%
\global\long\def\piv{v}%

\global\long\def\ib{i}%
\global\long\def\is{\alpha}%

\global\long\def\pbndo{\Gamma}%
\global\long\def\bndoo{\pbndo_{0}}%
 
\global\long\def\bndot{\pbndo_{t}}%
\global\long\def\intb{\inter{\body}}%
\global\long\def\bndb{\bdry\body}%

\global\long\def\cloo{\cl{\gO}}%

\global\long\def\nor{\nu}%
\global\long\def\Nor{\mathbf{N}}%

\global\long\def\dA{\dee A}%

\global\long\def\dV{\dee V}%

\global\long\def\eps{\varepsilon}%

\global\long\def\tv{v}%
\global\long\def\av{u}%

\global\long\def\svs{\mathcal{W}}%
\global\long\def\vs{\mathbf{V}}%
\global\long\def\avs{\mathbf{U}}%
\global\long\def\affsp{\mathcal{A}}%
\global\long\def\man{\mathcal{M}}%
\global\long\def\odman{\mathcal{N}}%
\global\long\def\subman{\mathcal{V}}%
\global\long\def\pt{p}%

\global\long\def\vbase{e}%
\global\long\def\sbase{\v e}%
\global\long\def\msbase{\mathfrak{e}}%
\global\long\def\vect{v}%
\global\long\def\dbase{\sbase}%

\global\long\def\chart{\varphi}%
\global\long\def\Chart{\Phi}%

\global\long\def\mind{\alpha}%
\global\long\def\vb{W}%
\global\long\def\vbp{\pi}%

\global\long\def\vbt{\mathcal{E}}%
\global\long\def\fib{\vs}%
\global\long\def\vbts{W}%
\global\long\def\avb{U}%
\global\long\def\vbp{\xi}%

\global\long\def\chart{\vph}%
\global\long\def\vbchart{\Phi}%

\global\long\def\jetb#1{J^{#1}}%
\global\long\def\jet#1{j^{1}(#1)}%
\global\long\def\tjet{\tilde{\jmath}}%

\global\long\def\Jet#1{J^{1}(#1)}%

\global\long\def\jetm{j}%

\global\long\def\coj{\mathfrak{d}}%

\global\long\def\alt{\mathfrak{A}}%

\global\long\def\pou{\eta}%

\global\long\def\ext{{\textstyle \bigwedge}}%
\global\long\def\forms{\Omega}%

\global\long\def\dotwedge{\dot{\mbox{\ensuremath{\wedge}}}}%

\global\long\def\vel{\theta}%

\global\long\def\Jac{\mathcal{J}}%

\global\long\def\contr{\mathbin{\raisebox{0.4pt}{\mbox{\ensuremath{\lrcorner}}}}}%
\global\long\def\fcor{\llcorner}%
\global\long\def\bcor{\lrcorner}%
\global\long\def\fcontr{\mathbin{\raisebox{0.4pt}{\mbox{\ensuremath{\llcorner}}}}}%

\global\long\def\lie{\mathcal{L}}%

\global\long\def\ssym#1#2{\ext^{#1}T^{*}#2}%

\global\long\def\sh{^{\sharp}}%

\global\long\def\nfo{\ext^{n}T^{*}\base}%
\global\long\def\dfs{\ext^{d}T^{*}\base}%
\global\long\def\dmfs{\ext^{d-1}T^{*}\base}%

\global\long\def\spc{\mathcal{S}}%
\global\long\def\sptm{\mathcal{E}}%
\global\long\def\evnt{e}%
\global\long\def\frame{\Psi}%

\global\long\def\timeman{\mathcal{T}}%
\global\long\def\zman{t}%
\global\long\def\dims{n}%
\global\long\def\m{\dims-1}%
\global\long\def\dimw{m}%

\global\long\def\wc{z}%

\global\long\def\fourv#1{\mbox{\ensuremath{\mathfrak{#1}}}}%

\global\long\def\body{\mathcal{B}}%
\global\long\def\man{\mathcal{M}}%
\global\long\def\var{\mathcal{V}}%
\global\long\def\base{\mathcal{X}}%
\global\long\def\fb{\mathcal{Y}}%
\global\long\def\srfc{\mathcal{Z}}%
\global\long\def\dimb{n}%
\global\long\def\dimf{m}%
\global\long\def\afb{\mathcal{Z}}%

\global\long\def\bdry{\partial}%

\global\long\def\gO{\varOmega}%

\global\long\def\reg{\gO}%
\global\long\def\bdrr{\bdry\reg}%

\global\long\def\bdom{\bdry\gO}%

\global\long\def\bndo{\partial\gO}%

\global\long\def\tpr{\vartheta}%

\global\long\def\mot{M}%
\global\long\def\vf{w}%
\global\long\def\const{h}%

\global\long\def\avf{u}%

\global\long\def\stn{\varepsilon}%
\global\long\def\djet{\chi}%

\global\long\def\jvf{\eps}%

\global\long\def\rig{r}%

\global\long\def\rigs{\mathcal{R}}%

\global\long\def\qrigs{\!/\!\rigs}%

\global\long\def\qd{\!/\,\!\kernel\diffop}%

\global\long\def\dis{\chi}%
\global\long\def\conf{\kappa}%
\global\long\def\invc{\hat{\conf}^{-1}}%
\global\long\def\dinvc{\hat{\conf}^{-1*}}%
\global\long\def\csp{\mathcal{Q}}%

\global\long\def\embds{\textrm{Emb}}%

\global\long\def\lc{A}%

\global\long\def\lv{\dot{A}}%
\global\long\def\alv{\dot{B}}%

\global\long\def\j{\mathop{\mathrm{j}}}%
\global\long\def\mapp{M}%
\global\long\def\J{J}%
\global\long\def\jex{\mathop{}\!\mathrm{j}}%

\global\long\def\fc{F}%
\global\long\def\load{f}%
\global\long\def\afc{g}%

\global\long\def\bfc{\mathbf{b}}%
\global\long\def\bfcc{b}%

\global\long\def\sfc{\mathbf{t}}%
\global\long\def\sfcc{t}%

\global\long\def\stm{\varsigma}%
\global\long\def\std{S}%
\global\long\def\tst{\sigma}%
\global\long\def\tstd{s}%
\global\long\def\st{\sigma}%
\global\long\def\vst{\varsigma}%
\global\long\def\vstd{S}%
\global\long\def\tstm{\sigma}%
\global\long\def\vstm{\varsigma}%

\global\long\def\stp{S_{P}}%
\global\long\def\slf{R}%

\global\long\def\crel{\Phi}%

\global\long\def\stmat{\tau}%

\global\long\def\gdiv{\bdry\textrm{iv\,}}%
\global\long\def\extjet{\mathfrak{d}}%

\global\long\def\smc#1{\mathfrak{#1}}%

\global\long\def\nhs{P}%
\global\long\def\nhsa{P}%
\global\long\def\nhsb{\underline{P}}%

\global\long\def\soc{Z}%

\global\long\def\sts{\varSigma}%
\global\long\def\spstd{\mathfrak{S}}%
\global\long\def\sptst{\mathfrak{T}}%
\global\long\def\spnhs{\mathcal{P}}%
\global\long\def\Ljj{\L{J^{1}(J^{k-1}\vb),\ext^{n}T^{*}\base}}%

\global\long\def\spsb{\text{{\Large \ensuremath{\Delta}}}}%

\global\long\def\ened{\mathfrak{w}}%
\global\long\def\energy{\mathfrak{W}}%

\global\long\def\ebdfc{T}%
\global\long\def\optimum{\st^{\textrm{opt}}}%
\global\long\def\scf{K}%

\global\long\def\grp{G}%
\global\long\def\gact{A}%
\global\long\def\gid{e}%
\global\long\def\gel{\ggm}%

\global\long\def\ael{\upsilon}%
\global\long\def\lal{\mathfrak{g}}%

\global\long\def\expr{\Pi}%

\global\long\def\aprop{Q}%

\global\long\def\flux{\omega}%
\global\long\def\aflux{\psi}%

\global\long\def\fform{\tau}%

\global\long\def\dimn{n}%

\global\long\def\sdim{{\dimn-1}}%

\global\long\def\fdens{\phi}%

\global\long\def\pform{s}%
\global\long\def\vform{\beta}%
\global\long\def\sform{\tau}%
\global\long\def\flow{\vf}%
\global\long\def\n{\m}%
\global\long\def\cmap{\mathfrak{t}}%
\global\long\def\vcmap{\varSigma}%

\global\long\def\mvec{\mathfrak{v}}%
\global\long\def\mveco#1{\mathfrak{#1}}%
\global\long\def\mv#1{\mathfrak{#1}}%
\global\long\def\smbase{\mathfrak{e}}%
\global\long\def\spx{\simp}%
\global\long\def\il{l}%
\global\long\def\awe{\frown}%

\global\long\def\hp{H}%
\global\long\def\ohp{h}%

\global\long\def\hps{G_{\dims-1}(T\spc)}%
\global\long\def\ohps{G^{\perp}_{\dims-1}(T\spc)}%

\global\long\def\hyper{\mathcal{S}}%

\global\long\def\hpsx{G_{\dims-1}(\tspc)}%
\global\long\def\ohpsx{G^{\perp}_{\dims-1}(\tspc)}%

\global\long\def\fbun{F}%

\global\long\def\flowm{\Phi}%

\global\long\def\tgb{T\spc}%
\global\long\def\ctgb{T^{*}\spc}%
\global\long\def\tspc{T_{\pis}\spc}%
\global\long\def\dspc{T^{*}_{\pis}\spc}%

\global\long\def\fflow{\fourv J}%
\global\long\def\fvform{\mathfrak{b}}%
\global\long\def\fsform{\mathfrak{t}}%
\global\long\def\fpform{\mathfrak{s}}%
\global\long\def\lfc{\mathfrak{F}}%

\global\long\def\maxw{\mathfrak{g}}%
\global\long\def\frdy{\mathfrak{f}}%
\global\long\def\ptnl{\gf}%
\global\long\def\pts{\Psi}%
\global\long\def\tptn{\Psi}%
\global\long\def\vptn{\mathfrak{a}}%
\global\long\def\mtst{\tstd_{M}}%
\global\long\def\mvst{\vstd_{M}}%

\global\long\def\sobp#1#2{W^{#1}_{#2}}%

\global\long\def\inner#1#2{\left\langle #1,#2\right\rangle }%

\global\long\def\fields{\sobp pk(\vb)}%

\global\long\def\bodyfields{\sobp p{k_{\partial}}(\vb)}%

\global\long\def\forces{\sobp pk(\vb)^{*}}%

\global\long\def\bfields{\sobp p{k_{\partial}}(\vb\resto{\bndo})}%

\global\long\def\loadp{(\sfc,\bfc)}%

\global\long\def\strains{\lp p(\jetb k(\vb))}%

\global\long\def\stresses{\lp{p'}(\jetb k(\vb)^{*})}%

\global\long\def\diffop{D}%

\global\long\def\strainm{E}%

\global\long\def\incomps{\vbts_{\yieldf}}%

\global\long\def\devs{L^{p'}(\eta^{*}_{1})}%

\global\long\def\incompsns{L^{p}(\eta_{1})}%

\global\long\def\testf{\mathcal{D}}%
\global\long\def\dists{\mathcal{D}'}%

\global\long\def\codiv{\boldsymbol{\partial}}%

\global\long\def\currof#1{\tilde{#1}}%

\global\long\def\chn{c}%
\global\long\def\chnsp{\mathbf{C}}%

\global\long\def\current{T}%
\global\long\def\curr{R}%

\global\long\def\curd{S}%
\global\long\def\curwd#1{\wh{#1}}%
\global\long\def\curnd#1{\wh{#1}}%

\global\long\def\contrf{{\scriptstyle \smallfrown}}%

\global\long\def\prodf{{\scriptstyle \smallsmile}}%

\global\long\def\form{\omega}%

\global\long\def\dens{\rho}%

\global\long\def\simp{s}%
\global\long\def\ssimp{\Delta}%
\global\long\def\cpx{K}%

\global\long\def\cell{C}%

\global\long\def\chain{B}%
\global\long\def\A{A}%
\global\long\def\B{B}%

\global\long\def\ach{A}%

\global\long\def\coch{X}%

\global\long\def\scale{s}%

\global\long\def\fnorm#1{\norm{#1}^{\flat}}%

\global\long\def\chains{\mathcal{A}}%

\global\long\def\ivs{\boldsymbol{U}}%

\global\long\def\mvs{\boldsymbol{V}}%

\global\long\def\cvs{\boldsymbol{W}}%

\global\long\def\ndual#1{#1'}%

\global\long\def\nd{'}%

\global\long\def\cee#1{C^{#1}}%

\global\long\def\lone{\{L^{1}\}}%

\global\long\def\linf{L^{\infty}}%

\global\long\def\lp#1{L^{#1}}%

\global\long\def\ofbdo{(\bndo)}%

\global\long\def\ofclo{(\cloo)}%

\global\long\def\vono{(\gO,\rthree)}%

\global\long\def\lomu{\{L^{1,\mu}\}}%
\global\long\def\limu{L^{\infty,\mu}}%
\global\long\def\limub{\limu(\body,\rthree)}%
\global\long\def\lomub{\lomu(\body,\rthree)}%

\global\long\def\vonbdo{(\bndo,\rthree)}%
\global\long\def\vonbdoo{(\bndoo,\rthree)}%
\global\long\def\vonbdot{(\bndot,\rthree)}%

\global\long\def\vonclo{(\cl{\gO},\rthree)}%

\global\long\def\strono{(\gO,\reals^{6})}%

\global\long\def\sob{\{W^{1}_{1}\}}%

\global\long\def\sobb{\sob(\gO,\rthree)}%

\global\long\def\lob{\lone(\gO,\rthree)}%

\global\long\def\lib{\linf(\gO,\reals^{12})}%

\global\long\def\ofO{(\gO)}%

\global\long\def\oneo{{1,\gO}}%
\global\long\def\onebdo{{1,\bndo}}%
\global\long\def\info{{\infty,\gO}}%

\global\long\def\infclo{{\infty,\cloo}}%

\global\long\def\infbdo{{\infty,\bndo}}%
\global\long\def\lobdry{\lone(\bdry\gO,\rthree)}%

\global\long\def\ld{LD}%

\global\long\def\ldo{\ld\ofO}%
\global\long\def\ldoo{\ldo_{0}}%

\global\long\def\trace{\gamma}%
\global\long\def\dtrace{\delta}%
\global\long\def\gtrace{\beta}%

\global\long\def\pr{\proj_{\rigs}}%

\global\long\def\pq{\proj}%

\global\long\def\qr{\,/\,\reals}%

\global\long\def\aro{S_{1}}%
\global\long\def\art{S_{2}}%

\global\long\def\mo{m_{1}}%
\global\long\def\mt{m_{2}}%

\global\long\def\ebdfc{T}%

\global\long\def\mini{\Omega}%
\global\long\def\optimum{s^{\mathrm{opt}}}%
\global\long\def\scf{K}%
\global\long\def\opsf{\st^{\mathrm{opt}}}%
\global\long\def\doptimum{s^{\opt,{\scriptscriptstyle D}}}%
\global\long\def\loptimum{s^{\opt,{\scriptscriptstyle \mathcal{M}}}}%

\global\long\def\fsubs{M}%

\global\long\def\yieldc{B}%

\global\long\def\yieldf{Y}%

\global\long\def\trpr{\pi_{P}}%

\global\long\def\devpr{\pi_{\devsp}}%

\global\long\def\prsp{P}%

\global\long\def\devsp{D}%

\global\long\def\ynorm#1{\|#1\|_{\yieldf}}%

\global\long\def\colls{\Psi}%

\global\long\def\aro{S_{1}}%
\global\long\def\art{S_{2}}%

\global\long\def\mo{m_{1}}%
\global\long\def\mt{m_{2}}%

\global\long\def\trps{^{\mathsf{T}}}%

\global\long\def\hb{^{\mathrm{hb}}}%

\global\long\def\yieldst{s_{Y}}%

\global\long\def\yieldc{B}%

\global\long\def\lcap{C}%

\global\long\def\yieldf{Y}%

\global\long\def\sphpr{\pi_{P}}%

\global\long\def\devpr{\pi_{\devsp}}%

\global\long\def\prsp{P}%

\global\long\def\devsp{D}%

\global\long\def\ynorm#1{\|#1\|_{\yieldf}}%

\global\long\def\colls{\Psi}%

\global\long\def\cone{Q}%
\global\long\def\fpr{\Pi}%
\global\long\def\fprd{\fpr_{\devsp}}%
\global\long\def\fprp{\fpr_{\prsp}}%
\global\long\def\find{I_{\devsp}}%
\global\long\def\finp{I_{\prsp}}%
\global\long\def\fnorm#1{\norm{#1}_{\devsp}}%

\global\long\def\rig{r}%
\global\long\def\rigs{\mathcal{R}}%
\global\long\def\qrigs{\!/\!\rigs}%
\global\long\def\anv{\omega}%
\global\long\def\I{I}%
\global\long\def\mone{M_{1}}%

\global\long\def\bd{BD}%

\global\long\def\po{\proj_{0}}%
\global\long\def\normp#1{\norm{#1}'_{\ld}}%

\global\long\def\ssx{S}%

\global\long\def\smap{s}%

\global\long\def\smat{\chi}%

\global\long\def\sx{e}%

\global\long\def\snode{P}%
\global\long\def\newmacroname{\{\}}%

\global\long\def\elem{e}%

\global\long\def\nel{L}%

\global\long\def\el{l}%

\global\long\def\gr{g}%
\global\long\def\ngr{G}%

\global\long\def\eldof{\alpha}%

\global\long\def\glbs{\psi}%

\global\long\def\ipln{\phi}%

\global\long\def\ndof{D}%

\global\long\def\dof{d}%

\global\long\def\nldof{N}%

\global\long\def\ldof{n}%

\global\long\def\lvf{\chi}%

\global\long\def\amat{A}%
\global\long\def\bmat{B}%

\global\long\def\subsp{\mathcal{M}}%
\global\long\def\zerofn{Z}%

\global\long\def\snomat{E}%

\global\long\def\femat{E}%

\global\long\def\tmat{T}%

\global\long\def\fvec{f}%

\global\long\def\snsp{\mathcal{S}}%

\global\long\def\slnsp{\Phi}%
\global\long\def\dslnsp{\Phi^{{\scriptscriptstyle D}}}%

\global\long\def\ro{r_{1}}%

\global\long\def\rtwo{r_{2}}%

\global\long\def\rth{r_{3}}%

\global\long\def\fmax{M}%

\global\long\def\dform{\psi}%

\global\long\def\srfc{\mathcal{S}}%

\global\long\def\semib{\mathrm{SB}}%

\global\long\def\tm#1{\overrightarrow{#1}}%
\global\long\def\tmm#1{\underrightarrow{\overrightarrow{#1}}}%

\global\long\def\itm#1{\overleftarrow{#1}}%
\global\long\def\itmm#1{\underleftarrow{\overleftarrow{#1}}}%

\global\long\def\ptrac{\mathcal{P}}%

\global\long\def\nh#1{\hat{#1}}%
\global\long\def\nj{\hat{\jmath}}%
\global\long\def\nJ{\hat{J}}%
\global\long\def\rin#1{\mathfrak{#1}}%
\global\long\def\npi{\hat{\pi}}%
\global\long\def\rp{\rin p}%
\global\long\def\rq{\rin q}%
\global\long\def\rr{\rin r}%

\global\long\def\xty{(\base,\fb)}%
\global\long\def\xts{(\base,\spc)}%
\global\long\def\r{r}%
\global\long\def\ntm{(\reals^{n},\reals^{m})}%

\global\long\def\tproj{\frame_{\timeman}}%
\global\long\def\sproj{\frame_{\spc}}%

\global\long\def\cons{c}%
\global\long\def\optm{\go}%
\global\long\def\flxs{\mathcal{W}}%
\global\long\def\cost{Q}%

\global\long\def\mtn{e}%
\global\long\def\sppp{\lambda}%

\global\long\def\mtsp{\mathscr{E}}%

\global\long\def\disp{g}%
\global\long\def\diffs{G}%

\global\long\def\bv{BV}%

\global\long\def\Charge{Q}%

\global\long\def\pole{q}%

\global\long\def\pdens{\rho}%

\global\long\def\ms#1{\mathfrak{#1}}%

\global\long\def\Hfl{\Phi}%

\global\long\def\cud{\v j}%
\global\long\def\mag{\v M}%
\global\long\def\vp{\v A}%
\global\long\def\magf{\v B}%
\global\long\def\magi{\v H}%

\global\long\def\pfun{\mathscr{P}}%

\title[Electrostatics as a Stress Theory]{Electrostatics as a Stress Theory in $\rthree$: Free and Bound Charge,
Gauge Invariance, and Mechanical Interactions}
\author{Vladimir Gol'dshtein, Wolfgang H.~Müller, and Reuven Segev}
\date{Draft, October 2026}
\begin{abstract}
We formulate electrostatics in $\rthree$ as the simplest instance
of the theory of stresses: the virtual velocity field of continuum
mechanics is simply replaced by a scalar potential field. A charge
distribution is then a linear functional on potential fields whose
value is the potential energy, and it is represented by a volume charge
density and by surface charge densities on the boundaries of subbodies.
Cauchy's theorem gives a vector field $\v P$, the analog of the traction
stress, and a scalar field $r$, so that the energy is $\int(r\vph+\v P\cdot\nabla\vph)\dV$.
The pair $(r,\v P)$, the analog of a variational stress, splits every
charge distribution uniquely into free charge, with density $r$,
and a dipole distribution, with density $\v P$, which carries the
bound charges $-\nabla\cdot\v P$ and $\v P\cdot\v{\nor}$; the fields
$r$ and $\v P$ are the densities of charge and of dipole moment.
We do not impose gauge invariance of the energy. Instead, we show
that it is the analog of the balance of forces on every subbody, and
that it singles out the neutral charge distributions. For charges
embedded in matter, the forces exerted by the electric field are derived.
They are shown to be gauge invariant for every charge distribution,
and they include an asymmetric active stress $\v E\otimes\v P$ whose
skew-symmetric part is the couple density $\v P\cross\v E$.
\end{abstract}

\maketitle
\tableofcontents{}

\section{Introduction}

Standard treatments of the electrostatics of ponderable bodies (e.g.,
{[}Jac99, Zan13{]}) introduce a polarization field, split the charge
into free and bound parts, and then postulate the body force and the
body couple acting on polarized matter. We take a different route.
We regard a charge distribution in a region $\gO\subset\rthree$ as
a linear functional acting on potential fields $\vph$, and we read
its value as the potential energy of the charge in the potential field.
Forces on matter are, in turn, linear functionals acting on virtual
velocity fields. We obtain them from the energy by assuming that the
charges are embedded in the material and that the system is conservative.
No constitutive relations are used.

The forces and couples that the electric field exerts on polarized
matter have a long history in continuum mechanics. Toupin {[}Tou56{]}
formulated the theory of the elastic dielectric, where the couple
$\v P\cross\v E$ exerted by the electric field on the polarized material
makes the stress in the material nonsymmetric. Maugin {[}Mau80{]}
applied the method of virtual power to electromagnetic interactions,
and comprehensive treatments can be found in {[}EM90, Kov00, DO14{]}.
We do not aim at new field equations or constitutive relations here.
What we offer is of a different nature: (i) we treat the charge distribution
itself as a force system in the sense of stress theory, so that the
split into free and bound charge follows from Cauchy's theorem instead
of being postulated through a polarization field; (ii) we show that
gauge invariance of the energy is the analog of the balance of forces,
that is, a way of classifying charge systems rather than a postulate;
and (iii) the forces on embedded charges, including the asymmetric
stress and the couple $\v P\cross\v E$, follow from the energy by
a single differentiation, and they are gauge invariant for every charge
system.

The guiding observation is that, seen this way, electrostatics is
a trivial example of the theory of stresses in $\rthree$: we get
it from continuum mechanics simply by replacing the virtual velocity
field with a scalar potential field. In this analogy: 
\begin{itemize}
\item the charge distribution plays the role of a force system, with volume
charge as body force and surface charge as surface traction; 
\item a vector field $\v P$ plays the role of the stress, and a scalar
field $r$ plays the role of the unbalanced body force; 
\item the potential energy plays the role of the virtual power; 
\item gauge transformations $\vph\mapsto\vph+c$, where $c$ is a constant,
play the role of rigid translations. 
\end{itemize}
Cauchy's theorem then gives the general representation of the energy,
\begin{equation}
Q_{\gO}(\vph)=\int_{\gO}\big(r\vph+\v P\cdot\nabla\vph\big)\dV=\int_{\gO}r\vph\dV-\int_{\gO}\v P\cdot\v E\dV,\label{eq:genU}
\end{equation}
the analog of the representation of a force system by a variational
stress {[}Seg02{]}, {[}Seg23, Ch.~10{]}. We show that the pair $(r,\v P)$
decomposes the charge uniquely into free charge, with density $r$,
and a distribution of dipoles, with density $\v P$. No relation between
the two is assumed.

In particular, we do not postulate gauge invariance of the energy.
If $\vph$ is a variation of the potential, a constant in a bounded
region $\gO\subset\rthree$ is an admissible variation, and the energy
of a body carrying a net charge $Q$ changes by $cQ$. Requiring invariance
would simply rule out net charge by assumption. What we find instead
is that requiring the energy of every subbody to be gauge invariant
is the exact analog of requiring the forces acting on every subbody
to add up to zero. In continuum mechanics, the latter singles out
the equilibrated force systems, those represented by the stress acting
on the velocity gradient alone, $\int\sigma_{ij}w_{i,j}\dV$. In electrostatics,
the former singles out the neutral charge systems, those with $r=0$,
for which the energy is 
\begin{equation}
U_{\gO}(\vph)=\int_{\gO}\v P\cdot\nabla\vph\dV=-\int_{\gO}\v P\cdot\v E\dV,\label{eq:newU}
\end{equation}
so that all the charge is bound in dipoles. Not every force system
is in equilibrium, and, likewise, not every charge system is neutral.
What does hold in general is that the mechanical forces exerted by
the electric field on matter are gauge invariant.

We start with the point charge and the point dipole, which motivate
the construction. In Section~\ref{sec:dipole} we compute the energy
of a dipole, the force and couple acting on it, and the virtual power
for a dipole embedded in a moving medium, first from the naive model
of two opposite charges and then in terms of linear functionals. Section~\ref{sec:mech}
reviews what we need from the theory of stresses in $\rthree$, including
the characterization of equilibrated force systems. In Section~\ref{sec:scalar}
we carry the theory over to scalar potentials and introduce the split
into free and bound charge. Section~\ref{sec:gauge} characterizes
the neutral charge systems by gauge invariance and also looks at the
singular field of a point charge. In Section~\ref{sec:forces} we
derive the mechanical forces exerted by the electric field on matter
in which the charges are embedded and show that they are gauge invariant,
and Section~\ref{sec:examples} gives examples. We close with some
remarks in Section~\ref{sec:remarks}.

Throughout, a comma denotes partial differentiation, we use the summation
convention, and the electric field is defined by $\v E=-\nabla\vph$,
so that $E_{i,j}=E_{j,i}$. We denote by $\v P$ the field that plays
the role of the polarization; the electric displacement $\v D$ appears
only in Section~\ref{sec:remarks}.

\section{Motivation: a single dipole}

\label{sec:dipole}

\subsection{The naive point of view}

In the naive picture, a dipole is the limit of two opposite point
charges $q$ and $-q$, where the small vector $d\v y$ points from
$-q$ to $q$ (Figure~\ref{fig:dipole}). We let the charge $q$
tend to infinity and $d\v y\to\v 0$ so that the dipole moment 
\begin{equation}
\v d:=q\,d\v y\label{eq:d}
\end{equation}
remains a well-defined finite vector.

\begin{figure}[ht]
\begin{tikzpicture}[scale=1.1]
\node[negchg] (m) at (0,0) {};
\node[poschg] (p) at (3.2,0.7) {};
\draw[vec] (m) -- (p) node[midway,above] {$d\v{y}$};
\node[below=2pt] at (m) {$-q$};
\node[right=2pt] at (p) {$q$};
\end{tikzpicture} \caption{A dipole.}
\label{fig:dipole} 
\end{figure}

\textbf{Energy.} When the dipole is placed in an electrostatic potential
field $\vph$, its potential energy is (Figure~\ref{fig:energy})
\begin{equation}
U_{\v d}=q\vph(\v y+d\v y)-q\vph(\v y)\cong q\vph_{,i}(\v y)\,dy_{i}=d_{i}\vph_{,i}(\v y),\label{eq:Ud}
\end{equation}
that is, 
\begin{equation}
U_{\v d}=\v d\cdot\nabla\vph(\v y)=-\v d\cdot\v E(\v y).
\end{equation}

\begin{figure}[ht]
\begin{tikzpicture}[scale=1.1]
\node[negchg] (m) at (0,0) {};
\node[poschg] (p) at (3.2,0.7) {};
\draw[vec] (m) -- (p) node[midway,above] {$d\v{y}$};
\node[left=2pt] at (m) {$-q$};
\node[above=2pt] at (p) {$q$};
\node[red!70!black,below=4pt] at (m) {$-q\vph(\v{y})$};
\node[red!70!black,below right=2pt and 2pt] at (p) {$q\vph(\v{y}+d\v{y})$};
\end{tikzpicture} \caption{A dipole in an electrostatic potential field.}
\label{fig:energy} 
\end{figure}

\textbf{Force.} The total force acting on the dipole is (Figure~\ref{fig:force})
\begin{equation}
f_{\v d\,i}=qE_{i}(\v y+d\v y)-qE_{i}(\v y)\cong qE_{i,j}\,dy_{j}=E_{i,j}d_{j},\label{eq:fd}
\end{equation}
that is, $\v f_{\v d}=(\nabla\v E)\,\v d=(\v d\cdot\nabla)\v E$.
In a uniform field the forces on the two charges cancel.

\begin{figure}[ht]
\begin{tikzpicture}[scale=1.1]
\node[negchg] (m) at (0,0) {};
\node[poschg] (p) at (3.2,0.7) {};
\draw[vec] (m) -- (p) node[midway,above] {$d\v{y}$};
\node[below right=1pt and 0pt] at (m) {$-q$};
\node[above=2pt] at (p) {$q$};
\draw[frc] (m) -- ++(-1.6,0.35) node[above] {$-q\v{E}(\v{y})$};
\draw[frc] (p) -- ++(1.4,-0.9) node[below right] {$q\v{E}(\v{y}+d\v{y})$};
\end{tikzpicture} \caption{A dipole in an electric field.}
\label{fig:force} 
\end{figure}

\textbf{Virtual power for an embedded dipole.} Now suppose that the
dipole is embedded in a material body, that is, its two charges are
attached to two neighboring material points, and that the body is
in virtual motion with velocity field $\v w$ (Figure~\ref{fig:embedded}).
The virtual power of the forces exerted by the electric field is 
\begin{equation}
\begin{split}W_{\v d} & =q\v E(\v y+d\v y)\cdot\v w(\v y+d\v y)-q\v E(\v y)\cdot\v w(\v y)\\
 & \cong E_{i,j}(\v y)\,(q\,dy_{j})\,w_{i}(\v y)+E_{i}(\v y)\,(q\,dy_{j})\,w_{i,j}(\v y),
\end{split}
\label{eq:Wd}
\end{equation}
and in the limit 
\begin{equation}
W_{\v d}=E_{i,j}d_{j}w_{i}+E_{i}d_{j}w_{i,j}=\v f_{\v d}\cdot\v w+(\v E\otimes\v d):\nabla\v w,\label{eq:Wd2}
\end{equation}
all fields evaluated at $\v y$.

\begin{figure}[ht]
\begin{tikzpicture}[scale=1.1]
\node[negchg] (m) at (0,0) {};
\node[poschg] (p) at (3.2,0.7) {};
\draw[vec] (m) -- (p) node[midway,above] {$d\v{y}$};
\node[below right=1pt and 0pt] at (m) {$-q$};
\node[right=3pt] at (p) {$q$};
\draw[frc] (m) -- ++(-1.6,0.35) node[above] {$-q\v{E}(\v{y})$};
\draw[frc] (p) -- ++(1.4,-0.9) node[below right] {$q\v{E}(\v{y}+d\v{y})$};
\draw[vel] (m) -- ++(0.9,1.0) node[above] {$\v{w}(\v{y})$};
\draw[vel] (p) -- ++(0.7,1.25) node[above] {$\v{w}(\v{y}+d\v{y})$};
\end{tikzpicture} \caption{An embedded dipole in virtual motion.}
\label{fig:embedded} 
\end{figure}

For a uniform velocity field only the first term survives, and it
gives the power of the force $\v f_{\v d}$. For a rigid rotation
$\v w=\v{\go}\cross\v x$, we have $w_{i,j}=\eps_{ikj}\omega_{k}$
and 
\begin{equation}
E_{i}d_{j}w_{i,j}=\v{\go}\cdot(\v d\cross\v E),
\end{equation}
so the skew-symmetric part of the tensor $\v E\otimes\v d$ represents
the couple $\v d\cross\v E$. In general, if we split $w_{i,j}$ into
its symmetric and skew-symmetric parts, the symmetric part of $E_{i}d_{j}$
acts on the rate of deformation. Since $E_{i}d_{j}$ acts on the velocity
gradient to produce power, it is a stress, concentrated at $\v y$
and asymmetric. So the expression (\ref{eq:Wd2}) tells us considerably
more than the total force $\v f_{\v d}$ does.

\subsection{The dipole as a linear functional}

With (\ref{eq:Ud}) in mind, we drop the limit of two charges and
define a dipole at $\v y$ as the linear functional on potential fields
\begin{equation}
Q_{\v d}(\vph):=\v d\cdot\nabla\vph(\v y),\label{eq:Qd}
\end{equation}
which we read as the potential energy of the dipole in the potential
field $\vph$. Compare this with the point charge $q$ at $\v y$,
the functional 
\begin{equation}
Q_{q}(\vph):=q\,\vph(\v y).\label{eq:Qq}
\end{equation}
Two properties of the dipole functional, which the point charge does
not share, will be central in what follows. 
\begin{enumerate}
\item[(a)] \emph{A dipole is neutral:} $Q_{\v d}(1)=0$. Equivalently, its energy
is unchanged when a constant is added to the potential, $Q_{\v d}(\vph+c)=Q_{\v d}(\vph)$;
it is gauge invariant. For a point charge, $Q_{q}(\vph+c)=Q_{q}(\vph)+cq$. 
\item[(b)] \emph{The energy depends on the potential only through the electric
field:} $Q_{\v d}(\vph)=-\v d\cdot\v E(\v y)$. 
\end{enumerate}
\textbf{The embedded dipole.} Let the dipole be embedded in a body
undergoing a smooth virtual motion $\dis:\reals\cross\gO\to\rthree$
with $\dis_{0}=\mathrm{Id}$ and velocity $\v w=\bdry_{t}\dis_{t}|_{t=0}$.
Since $d\v y$ is a material line element, the dipole is carried by
the deformation gradient: at time $t$, the dipole is $\v d_{t}=\nabla\dis_{t}(\v y)\,\v d$,
located at $\dis_{t}(\v y)$. Its energy is 
\begin{equation}
U_{t}=\v d_{t}\cdot\nabla\vph(\dis_{t}(\v y))=\dis_{t\,i,j}(\v y)\,d_{j}\,\vph_{,i}(\dis_{t}(\v y)).
\end{equation}
Assuming that the system is conservative, i.e., that the virtual power
of the forces exerted by the electric field equals minus the rate
of change of the potential energy, and using $\dis_{0\,i,j}(\v y)=\delta_{ij}$,
$\bdry_{t}\dis_{t\,i,j}(\v y)|_{0}=w_{i,j}(\v y)$, and $\bdry_{t}\vph_{,i}(\dis_{t}(\v y))|_{0}=\vph_{,ik}(\v y)\,w_{k}(\v y)$,
\begin{equation}
F_{\v d}(\v w):=-\frac{d}{dt}U_{t}\Big|_{t=0}=-d_{j}\big(\vph_{,ji}w_{i}+\vph_{,i}w_{i,j}\big)=E_{i,j}d_{j}w_{i}+E_{i}d_{j}w_{i,j},\label{eq:Fd}
\end{equation}
in agreement with (\ref{eq:Wd2}), with all fields evaluated at $\v y$.
Recall that $\v E:=-\nabla\vph$. For a point charge carried by the
motion, $U_{t}=q\vph(\dis_{t}(\v y))$ and $F_{q}(\v w)=q\v E(\v y)\cdot\v w(\v y)$,
the Coulomb force. Although the energy of a point charge is not gauge
invariant, the force is: $U_{t}$ changes by the constant $cq$, which
does not depend on $t$.

\textbf{Toward distributions of charges and dipoles.} In the rest
of the paper we look for the continuum counterpart of (\ref{eq:Qq}),
(\ref{eq:Qd}) and (\ref{eq:Fd}). We show that every charge distribution,
localized to subbodies, is a superposition of point charges and point
dipoles, that the two parts can be separated uniquely, and that the
purely dipolar distributions are exactly those that are gauge invariant
on every subbody. To see why the latter is natural, we first recall
what the analogous condition does in continuum mechanics.

\section{Stresses and the balance of forces in continuum mechanics}

\label{sec:mech}

We begin by recalling the parts of stress theory that we need for
the analogy. That the stress exists is Cauchy's classical theorem
{[}Nol59{]}, {[}TT60, Sect.~203, p.~543{]}. The distinction between
the traction stress $\sigma$ and the variational stress $(\v r,\sigma)$,
explained below, was introduced in {[}Seg02{]} for a general manifold,
where the traction stress is called the Cauchy stress; see also {[}Seg23,
Ch.~10{]}.

\textbf{Force systems.} For each regular subbody $\gO$ of a body,
a force $F_{\gO}$ is a linear functional on virtual velocity fields
$\v w$, represented by a body force $\v b$ and a surface force $\v t_{\gO}$
on $\bdry\gO$: 
\begin{equation}
F_{\gO}(\v w)=\int_{\gO}\v b\cdot\v w\dV+\int_{\bdry\gO}\v t_{\gO}\cdot\v w\dA.\label{eq:FOm}
\end{equation}
The value $F_{\gO}(\v w)$ is the virtual power.

\textbf{Cauchy's theorem: the traction stress.} Under the usual assumptions
(dependence of $\v t_{\gO}$ on $\gO$ only through the normal, and
boundedness) {[}Nol59, TT60{]}, there is a tensor field $\sigma$,
the (traction) stress, such that 
\begin{equation}
\v t_{\gO}=\sigma\v{\nor},\label{eq:cauchy}
\end{equation}
where $\v{\nor}$ is the outward unit normal. We refer to $\sigma$
as the \emph{traction stress}. It determines the surface force on
every surface in the body, in particular on internal cuts, and, conversely,
these surface forces determine it. The traction stress assigns to
the value $\v w(\v x)$ of a velocity field the vector $\sigma^{T}\v w$,
the power flux, whose flux through $\bdry\gO$ is the power of the
surface forces. So the power of the traction stress is 
\begin{equation}
\int_{\bdry\gO}\v t_{\gO}\cdot\v w\dA=\int_{\bdry\gO}(\sigma^{T}\v w)\cdot\v{\nor}\dA=\int_{\gO}\nabla\cdot(\sigma^{T}\v w)\dV=\int_{\gO}\big((\nabla\cdot\sigma)\cdot\v w+\sigma_{ij}w_{i,j}\big)\dV,\label{eq:tpower}
\end{equation}
where $(\nabla\cdot\sigma)_{i}=\sigma_{ij,j}$, and 
\begin{equation}
F_{\gO}(\v w)=\int_{\gO}\v b\cdot\v w\dV+\int_{\gO}\nabla\cdot(\sigma^{T}\v w)\dV.\label{eq:Ftrac}
\end{equation}
Note that the traction stress acts only on the values of $\v w$,
and that its power on a subbody is a boundary integral.\footnote{In the geometric setting, $\sigma(\v w)$ is an $(n-1)$-form and
(\ref{eq:Ftrac}) reads $F_{\gO}(\v w)=\int_{\gO}\big(\v b(\v w)+d(\sigma(\v w))\big)$
{[}Seg02, Sect.~IV{]}.} The traction stress alone does not determine the force system: the
body force $\v b$ is additional, independent data.

\textbf{The variational stress.} Expanding the divergence in (\ref{eq:Ftrac})
gives 
\begin{equation}
F_{\gO}(\v w)=\int_{\gO}\big(\v r\cdot\v w+\sigma_{ij}w_{i,j}\big)\dV,\qquad\v r:=\v b+\nabla\cdot\sigma.\label{eq:mechvar}
\end{equation}
The pair $S=(\v r,\sigma)$ is the \emph{variational stress field}
{[}Seg02, Sect.~V{]}, {[}Seg23, Sect.~10.7{]}. At each point, it
acts on the first jet of $\v w$, i.e., on the pair $(\v w,\nabla\v w)$,
and the integrand $\v r\cdot\v w+\sigma_{ij}w_{i,j}$ is a density
of power. Three of its features are worth stressing. 
\begin{enumerate}
\item[(a)] It represents the whole force system, body force and surface forces,
on all subbodies by a single field that does not depend on the subbody:
we get $F_{\gO}$ just by integrating over $\gO$, with no boundary
term. The surface forces come back only through the divergence theorem. 
\item[(b)] It arises directly when we regard forces as continuous linear functionals
on spaces of differentiable velocity fields, the tangent vectors to
the configuration space of the body. By a representation theorem,
every such functional is represented by a variational stress {[}Seg02,
Sect.~V{]}, {[}Seg23, Ch.~10{]}; in the non-smooth setting, the
variational stress may be a measure {[}Seg23, Ch.~21{]}. Neither
surface forces nor Cauchy's postulates are needed for this. 
\item[(c)] On a general manifold, there is no natural splitting of the jet of
$\v w$ into a value and a derivative, so that a variational stress
is not a pair of fields. In $\rthree$ the splitting is natural, and
$\v r$ and $\sigma$ are separately meaningful. 
\end{enumerate}
\textbf{The relation between the two.} The traction stress is the
part of the variational stress that acts on the gradient, $(\v r,\sigma)\mapsto\sigma$;
an invariant construction of this mapping on manifolds is given in
{[}Seg02, Sect.~VI{]} and {[}Seg23, Ch.~10{]}. The variational stress
also determines the body force, $\v b=\v r-\nabla\cdot\sigma$. Conversely,
the traction stress and the body force together determine the variational
stress, $S=(\v b+\nabla\cdot\sigma,\sigma)$. So a variational stress
carries the same information as a traction stress together with a
body force, but not the same information as the traction stress alone:
a given traction stress corresponds to a whole family of variational
stresses $(\v r,\sigma)$, parametrized by the arbitrary field $\v r$.

The two variants play different roles. The traction stress describes
the interactions across surfaces; it is what the Cauchy approach works
with, and what one measures on cuts. The variational stress describes
the density of power in the bulk, dual to the generalized velocity;
it is what the variational approach works with, and it is the one
that generalizes to manifolds and to higher-order theories. The traction
stress alone determines the variational stress only when we know that
$\v r=\v 0$, which is the case we characterize next.

\textbf{Rigid translations and the balance of forces.} A rigid translation
is a uniform velocity field $\v w=\v c$, i.e., an element of the
kernel of $\v w\mapsto\nabla\v w$. For a uniform field, $F_{\gO}(\v c)=\v c\cdot\big(\int_{\gO}\v b\dV+\int_{\bdry\gO}\v t_{\gO}\dA\big)$,
that is, $\v c$ dotted with the sum of the forces acting on $\gO$.
\begin{prop}
\label{prop:mechbalance} For a force system with continuous $\v b$
and $C^{1}$ stress $\sigma$, the following are equivalent. 
\begin{enumerate}
\item[(i)] The sum of the forces on every subbody vanishes: $\int_{\gO}\v b\dV+\int_{\bdry\gO}\v t_{\gO}\dA=\v 0$
for every $\gO$. 
\item[(ii)] The virtual power is invariant under rigid translations: $F_{\gO}(\v w+\v c)=F_{\gO}(\v w)$
for every $\gO$, $\v w$, and constant $\v c$. 
\item[(iii)] $\v r=\v 0$, i.e., $\v b+\nabla\cdot\sigma=\v 0$. 
\item[(iv)] $F_{\gO}(\v w)=\int_{\gO}\sigma_{ij}w_{i,j}\dV$ for every $\gO$
and $\v w$. 
\end{enumerate}
\end{prop}

\begin{proof}
(i)$\Leftrightarrow$(ii) by linearity, since $F_{\gO}(\v c)$ is
$\v c$ dotted with the sum of the forces. By (\ref{eq:mechvar}),
$F_{\gO}(\v c)=\v c\cdot\int_{\gO}\v r\dV$; this vanishes for every
$\gO$ and $\v c$ if and only if $\v r=\v 0$, by continuity, giving
(ii)$\Leftrightarrow$(iii). Finally, (iii)$\Leftrightarrow$(iv)
by (\ref{eq:mechvar}). 
\end{proof}

So the balance of forces on every subbody is exactly the condition
under which the force system is represented by the stress acting on
the velocity gradient alone. Without it, the stress does not determine
the force system, since any $\v r$ may be added. If we also require
invariance under rigid rotations, we obtain the symmetry of $\sigma$;
this has no analog in the scalar case below.

\section{The scalar analog: charge distributions}

\label{sec:scalar}

We now replace the virtual velocity field $\v w$ by a smooth scalar
potential field $\vph$ on $\rthree$. We may think of $\vph$ either
as the potential itself or as a variation of it, assuming that it
can be varied while the charge distribution is held fixed.

\textbf{Charge systems.} For each regular subbody $\gO$, a charge
distribution is a linear functional $Q_{\gO}$ on potential fields,
whose value $Q_{\gO}(\vph)$ is the potential energy of the charge
in $\gO$ under $\vph$. It is represented by a volume charge density
$\rho$ and a surface charge density $s_{\gO}$ on $\bdry\gO$: 
\begin{equation}
Q_{\gO}(\vph)=\int_{\gO}\rho\vph\dV+\int_{\bdry\gO}s_{\gO}\vph\dA.\label{eq:QOm}
\end{equation}
The surface charge $s_{\gO}$ depends on the subbody: it is the charge
exposed on $\bdry\gO$ when $\gO$ is cut out of the medium.

\textbf{Cauchy's theorem.} Under the same assumptions as in Section~\ref{sec:mech},
now for a scalar-valued surface density, Cauchy's theorem, in its
special case for scalar fluxes, gives a vector field $\v P$ such
that 
\begin{equation}
s_{\gO}=\v P\cdot\v{\nor}.
\end{equation}
This is the ``traction stress'' of the scalar theory; since the
generalized velocity is a scalar, the stress is a vector rather than
a tensor. The traction stress assigns to the value of the potential
the vector $\vph\v P$, the energy flux, whose flux through $\bdry\gO$
is the energy of the surface charge. Thus the energy of the traction
stress, the analog of (\ref{eq:tpower}), is 
\begin{equation}
\int_{\bdry\gO}s_{\gO}\vph\dA=\int_{\bdry\gO}(\vph\v P)\cdot\v{\nor}\dA=\int_{\gO}\nabla\cdot(\vph\v P)\dV=\int_{\gO}\big(\vph\nabla\cdot\v P+\v P\cdot\nabla\vph\big)\dV,\label{eq:Dpower}
\end{equation}
and, as in (\ref{eq:Ftrac}), 
\begin{equation}
Q_{\gO}(\vph)=\int_{\gO}\rho\vph\dV+\int_{\gO}\nabla\cdot(\vph\v P)\dV.\label{eq:Qtrac}
\end{equation}
Expanding the divergence gives 
\begin{equation}
Q_{\gO}(\vph)=\int_{\gO}\big(r\vph+\v P\cdot\nabla\vph\big)\dV,\qquad r:=\rho+\nabla\cdot\v P.\label{eq:var}
\end{equation}
The pair $(r,\v P)$ is the variational stress of the scalar theory;
it acts on the first jet $(\vph,\nabla\vph)$ of the potential, and
$r\vph+\v P\cdot\nabla\vph$ is a density of energy. As in Section~\ref{sec:mech},
the traction stress $\v P$ alone determines the surface charges but
not the charge system, while the variational stress $(r,\v P)$ determines
both the volume charge $\rho=r-\nabla\cdot\v P$ and the surface charges,
and conversely. In the mechanical analogy, $r$ corresponds to $\v r=\v b+\nabla\cdot\sigma$,
the part of the force system not balanced by the stress.

\textbf{Free and bound charge.} The pair $(r,\v P)$ is uniquely determined
by the charge system: $\v P$ by the surface charges on all subbodies,
through $s_{\gO}=\v P\cdot\v{\nor}$, and then $r$ by $r=\rho+\nabla\cdot\v P$.
Writing 
\begin{equation}
\rho=r+\rho_{b},\qquad\rho_{b}:=-\nabla\cdot\v P,\label{eq:decomp}
\end{equation}
the charge of every subbody splits uniquely into two parts. The first
is the free charge, with density $r$; its energy $\int_{\gO}r\vph\dV$
is the superposition of point charges (\ref{eq:Qq}) with $q=r\dV$.
The second is the charge bound in a distribution of dipoles with density
$\v P$, consisting of the bound volume charge $\rho_{b}$ and the
bound surface charge $s_{\gO}$; its energy $\int_{\gO}\v P\cdot\nabla\vph\dV$
is the superposition of point dipoles (\ref{eq:Qd}) with $\v d=\v P\dV$.
The decomposition is not postulated; it is the representation (\ref{eq:var})
given by Cauchy's theorem, and no relation between $r$ and $\v P$
is assumed. In the terminology of {[}Jac99, Zan13{]}, $r$ is the
free charge density $\rho_{f}$ and $\v P$ is the polarization.

\textbf{Uniqueness from the energies.} The pair $(r,\v P)$ is also
determined by the energies of the subbodies alone, without reference
to their surface charges. Let $(r,\v P)$ and $(r',\v P')$ be continuous
and represent the same charge system in the form (\ref{eq:var}),
so that 
\begin{equation}
\int_{\gO}\big((r-r')\vph+(\v P-\v P')\cdot\nabla\vph\big)\dV=0\label{eq:uniq}
\end{equation}
for every subbody $\gO$ and every potential $\vph$. With $\vph=1$,
$\int_{\gO}(r-r')\dV=0$ for every $\gO$. If $r-r'$ were positive
at a point, it would be positive on a ball about that point, and its
integral over the ball would be positive. Hence $r=r'$. With $\vph=x_{i}$,
(\ref{eq:uniq}) then gives $\int_{\gO}(P_{i}-P'_{i})\dV=0$ for every
$\gO$, and $\v P=\v P'$ in the same way.

The argument also gives $r$ and $\v P$ explicitly. Let $B_{\delta}$
be the ball of radius $\delta$ about a point $\v x_{0}$. Then $Q_{B_{\delta}}(1)=\int_{B_{\delta}}r\dV$
and 
\[
Q_{B_{\delta}}(x_{i}-x_{0i})=\int_{B_{\delta}}r\,(x_{i}-x_{0i})\dV+\int_{B_{\delta}}P_{i}\dV,
\]
where the first integral on the right is bounded by $\delta\,|B_{\delta}|\max_{B_{\delta}}|r|$.
Hence 
\begin{equation}
r(\v x_{0})=\lim_{\delta\to0}\frac{Q_{B_{\delta}}(1)}{|B_{\delta}|},\qquad P_{i}(\v x_{0})=\lim_{\delta\to0}\frac{Q_{B_{\delta}}(x_{i}-x_{0i})}{|B_{\delta}|}.\label{eq:moments}
\end{equation}
In words, the constant potential measures the charge, and the linear
potentials $x_{i}-x_{0i}$ measure the dipole moment about $\v x_{0}$:
for the point charge (\ref{eq:Qq}) they give $q$ and $q(\v y-\v x_{0})$,
and for the point dipole (\ref{eq:Qd}) they give $0$ and $\v d$.
So $r$ is the density of charge and $\v P$ is the density of dipole
moment. The potentials we use here are not compactly supported in
$\gO$, and this matters: the energy of a subbody has to be defined
for potentials that do not vanish on its boundary. On compactly supported
potentials alone, the densities are not determined: by (\ref{eq:freeDf})
below, the free charge $r=\nabla\cdot\v D_{f}$ and the dipole density
$-\v D_{f}$ have the same energy for every compactly supported potential
(cf.\ Remark~\ref{rem:wholespace}).

\textbf{The analogy.} Table~\ref{tab:analogy} summarizes the correspondence
with Section~\ref{sec:mech}. This is the sense in which electrostatics
is a trivial example of stresses in $\rthree$: the vector bundle
of generalized velocities is the trivial line bundle, Cauchy's theorem
for stresses reduces to Cauchy's theorem for scalar fluxes, and the
only ``rigid'' fields are constants.

\begin{table}[ht]
\global\long\def\arraystretch{1.25}%
\begin{tabular}{>{\raggedright}p{0.42\textwidth}>{\raggedright}p{0.42\textwidth}}
\hline 
\textbf{Continuum mechanics}  & \textbf{Electrostatics} \tabularnewline
\hline 
virtual velocity $\v w$  & potential $\vph$ \tabularnewline
velocity gradient $\nabla\v w$  & $\nabla\vph=-\v E$ \tabularnewline
rigid translations $\v w=\v c$  & gauge transformations $\vph\mapsto\vph+c$ \tabularnewline
body force $\v b$  & volume charge $\rho$ \tabularnewline
unbalanced force $\v r=\v b+\nabla\cdot\sigma$  & free charge $r=\rho+\nabla\cdot\v P$ \tabularnewline
surface traction $\v t_{\gO}=\sigma\v{\nor}$  & surface charge $s_{\gO}=\v P\cdot\v{\nor}$ \tabularnewline
traction stress $\sigma$  & $\v P$ \tabularnewline
power flux $\sigma^{T}\v w$  & energy flux $\vph\v P$ \tabularnewline
variational stress $(\v r,\sigma)$  & $(r,\v P)$ \tabularnewline
virtual power $F_{\gO}(\v w)$  & potential energy $Q_{\gO}(\vph)$ \tabularnewline
sum of forces on $\gO$  & total charge of $\gO$, $Q_{\gO}(1)$ \tabularnewline
stress power $\int\sigma_{ij}w_{i,j}\dV$  & $\int\v P\cdot\nabla\vph\dV$ \tabularnewline
balance $\v b+\nabla\cdot\sigma=\v 0$  & $\rho+\nabla\cdot\v P=0$ \tabularnewline
every subbody in equilibrium  & every subbody neutral \tabularnewline
\hline 
\end{tabular}\caption{The analogy between continuum mechanics and electrostatics.}
\label{tab:analogy} 
\end{table}

\section{Gauge invariance and neutral charge systems}

\label{sec:gauge}

\subsection{The main result}

Gauge invariance of the energy is not a general property of charge
systems. For a constant $c$, (\ref{eq:var}) gives 
\begin{equation}
Q_{\gO}(\vph+c)-Q_{\gO}(\vph)=c\,Q_{\gO}(1)=c\int_{\gO}r\dV,\label{eq:gaugechange}
\end{equation}
$c$ times the net charge of $\gO$. If $\vph$ is a variation of
the potential, this change is physical: shifting the potential of
a charged body does change its energy. So gauge invariance is a property
of a special class of charge systems, just as equilibrium is a property
of a special class of force systems. The following proposition characterizes
this class.
\begin{defn}
A charge system $\{Q_{\gO}\}$ is \emph{gauge invariant} if $Q_{\gO}(\vph+c)=Q_{\gO}(\vph)$
for every subbody $\gO$, every potential $\vph$, and some constant
$c\neq0$.
\end{defn}

\begin{prop}
\label{thm:main} For a charge system with continuous $\rho$ and
$C^{1}$ field $\v P$, the following are equivalent. 
\begin{enumerate}
\item[(i)] Every subbody is neutral: $\int_{\gO}\rho\dV+\int_{\bdry\gO}s_{\gO}\dA=0$
for every $\gO$. 
\item[(ii)] The charge system is gauge invariant. 
\item[(iii)] $r=0$, i.e., $\rho=-\nabla\cdot\v P$. 
\item[(iv)] $Q_{\gO}(\vph)=\int_{\gO}\v P\cdot\nabla\vph\dV=-\int_{\gO}\v P\cdot\v E\dV$
for every $\gO$ and $\vph$. 
\end{enumerate}
\end{prop}

\begin{proof}
The proof is word for word that of Proposition~\ref{prop:mechbalance}.
(i)$\Leftrightarrow$(ii) because $Q_{\gO}(c)=cQ_{\gO}(1)$ and $Q_{\gO}(1)$
is the total charge of $\gO$. By (\ref{eq:var}), $Q_{\gO}(1)=\int_{\gO}r\dV$,
which vanishes for every $\gO$ if and only if $r=0$. Finally, (iii)$\Leftrightarrow$(iv)
by (\ref{eq:var}). 
\end{proof}

So gauge invariance of the energy is to electrostatics what the balance
of forces on every subbody is to continuum mechanics: it singles out
the neutral charge systems, those without free charge, and the representation
(\ref{eq:newU}) of their energy is the analog of the representation
of the power of an equilibrated force system by $\int\sigma_{ij}w_{i,j}\dV$.
In both theories this is a classification, not a postulate. A body
carrying free charge is the analog of a body acted upon by unbalanced
forces.

\subsection{Consequences}

\textbf{The stress determines the energy of a neutral system.} In
general, $\v P$ determines only the bound charges, and the free charge
$r$ is an independent datum, just as $\v r$ is in (\ref{eq:mechvar}).
For a neutral system, $\v P$ determines the charge system completely:
\begin{equation}
\rho_{b}:=-\nabla\cdot\v P,\qquad s_{\gO}=\v P\cdot\v{\nor},\qquad U_{\gO}(\vph)=-\int_{\gO}\v P\cdot\v E\dV.\label{eq:bound}
\end{equation}
We write $\rho_{b}$ to indicate that this is the charge bound in
the dipole distribution.

\textbf{The energy depends only on the electric field.} For a neutral
system, the energy (\ref{eq:newU}) depends on $\vph$ only through
$\v E=-\nabla\vph$, as does the energy of a single dipole (property
(b) of Section~\ref{sec:dipole}). In general, the energy (\ref{eq:genU})
also depends on $\vph$ itself, through the free charge. The mechanical
forces of Section~\ref{sec:forces} nevertheless depend on $\v E$
only, for every charge system (Proposition~\ref{prop:forcegauge}).

\textbf{Neutral systems are dipole distributions.} A distribution
of dipoles (\ref{eq:Qd}) with density $\v P$, i.e., with dipole
moment $\v P\dV$ in the volume element $dV$, has, by superposition,
exactly the energy (\ref{eq:newU}). Each dipole is neutral, so such
a distribution is gauge invariant. Conversely, Proposition~\ref{thm:main}
shows that every gauge-invariant charge system with a smooth $\v P$
is of this form. Interpreting $\v P$ as the polarization, (\ref{eq:bound})
reproduces the bound charges $-\nabla\cdot\v P$ and $\v P\cdot\v{\nor}$
of {[}Jac99, Zan13{]}, with the correct signs.

The sign of the bound volume charge can be read off directly from
the model of Section~\ref{sec:dipole}. Take a chain of dipoles along
the $x_{1}$-axis whose moments increase with $x_{1}$. The positive
end of each dipole sits next to the negative end of the next, stronger
one, so a net negative charge, $-\bdry P_{1}/\bdry x_{1}$ per unit
volume, is left over.

\textbf{The meaning of the surface charge.} The surface charge $s_{\gO}=\v P\cdot\v{\nor}$
is defined for every subbody, not only for the boundary of the medium.
On an internal surface it is the charge exposed when the subbody is
cut out: the ends of the dipoles severed by the cut. This is the exact
analog of the traction on an internal surface in continuum mechanics,
which is the force exerted across the cut.

\subsection{A singular example: the point charge}

\label{sec:point}

Proposition~\ref{thm:main} assumes that $\v P$ is of class $C^{1}$.
The following example shows how the proposition extends to singular
fields, and what such fields represent. Let 
\begin{equation}
\v P(\v x)=\frac{q}{4\pi}\,\frac{\v x}{|\v x|^{3}},\qquad\v x\in\rthree\setminus\{\v 0\}.\label{eq:Dpoint}
\end{equation}
Then $\nabla\cdot\v P=0$ on $\rthree\setminus\{\v 0\}$, and $\int_{S}\v P\cdot\v{\nor}\dA=q$
for every closed surface $S$ enclosing the origin. Since $|\v P|=O(|\v x|^{-2})$
is locally integrable in $\rthree$, the energy $\int_{\gO}\v P\cdot\nabla\vph\dV$
is well defined for every regular region $\gO$ and every smooth $\vph$,
whether or not $\v 0\in\gO$.
\begin{prop}
\label{prop:point} Let $\v P$ be given by (\ref{eq:Dpoint}), and
let $\gO$ be a regular region with $\v 0\notin\bdry\gO$. Then 
\begin{equation}
U_{\gO}(\vph)=\int_{\gO}\v P\cdot\nabla\vph\dV=\begin{cases}
{\displaystyle \int_{\bdry\gO}(\v P\cdot\v{\nor})\,\vph\dA-q\,\vph(\v 0),} & \v 0\in\gO,\\[2ex]
{\displaystyle \int_{\bdry\gO}(\v P\cdot\v{\nor})\,\vph\dA,} & \v 0\notin\gO.
\end{cases}\label{eq:Upoint}
\end{equation}
Thus the charge system represented by $\v P$ consists of the surface
charge $s_{\gO}=\v P\cdot\v{\nor}$ on $\bdry\gO$ and, if $\v 0\in\gO$,
the bound point charge $-q$ at the origin, i.e., $\rho_{b}=-\nabla\cdot\v P=-q\delta_{\v 0}$
in the sense of distributions. Every such subbody is neutral, and
the energy is gauge invariant. 
\end{prop}

\begin{proof}
If $\v 0\notin\gO$, $\v P$ is smooth on $\overline{\gO}$ and the
divergence theorem applies directly, with $\nabla\cdot\v P=0$. If
$\v 0\in\gO$, let $B_{\eps}$ be a ball of radius $\eps$ about the
origin, contained in $\gO$. On $\gO\setminus B_{\eps}$, $\nabla\cdot(\vph\v P)=\v P\cdot\nabla\vph$,
and the outward normal on $\bdry B_{\eps}$, as part of the boundary
of $\gO\setminus B_{\eps}$, is $-\v x/|\v x|$. Hence 
\[
\int_{\gO\setminus B_{\eps}}\v P\cdot\nabla\vph\dV=\int_{\bdry\gO}(\v P\cdot\v{\nor})\,\vph\dA-\frac{q}{4\pi\eps^{2}}\int_{\bdry B_{\eps}}\vph\dA.
\]
As $\eps\to0$, the left side tends to $\int_{\gO}\v P\cdot\nabla\vph\dV$
by the integrability of $\v P$, and the last term tends to $q\vph(\v 0)$
by the continuity of $\vph$. Neutrality follows from (\ref{eq:Upoint})
with $\vph=1$: the surface charges total $q$ and cancel the point
charge $-q$. 
\end{proof}

Used as the field $\v P$ of a charge system, (\ref{eq:Dpoint}) therefore
does not represent an isolated point charge; it does, however, when
it is used to represent the free charge, as we show in Remark~\ref{rem:Df}
below. An isolated point charge $q$ at the origin is the charge system
with free charge $r=q\delta_{\v 0}$ and $\v P=\v 0$, whose energy
$q\vph(\v 0)$ is not gauge invariant. The field (\ref{eq:Dpoint})
represents a radial (``hedgehog'') polarization: the dipoles point
away from the origin, leaving the bound charge $-q$ at the center
and the compensating charge $+q$ spread over $\bdry\gO$. A bound
point charge $+q$ at the origin is obtained from $\v P=-q\v x/(4\pi|\v x|^{3})$.
\begin{rem}[The whole space]
\label{rem:wholespace} If $\gO=\rthree$ and the potentials are
compactly supported, there is no boundary term, and (\ref{eq:Upoint})
becomes 
\[
U(\vph)=\int_{\rthree}\v P\cdot\nabla\vph\dV=-q\,\vph(\v 0).
\]
The functional then acts exactly as an isolated point charge $-q$;
the compensating charge $+q$ has been sent to infinity, since $\int_{S_{R}}\v P\cdot\v{\nor}\dA=q$
for every sphere $S_{R}$ about the origin. Constants are not compactly
supported, so the gauge test is not available here, and there is no
contradiction. 
\end{rem}

\begin{rem}[Bound and free point charges]
\label{rem:free} Proposition~\ref{prop:point} and the decomposition
(\ref{eq:decomp}) clarify the status of point charges. A point charge
may enter a charge system in two ways: as free charge, $r=q\delta_{\v 0}$,
with no compensating charge and an energy that changes by $cq$ under
$\vph\mapsto\vph+c$; or as bound charge, through the singular field
$\v P=-q\v x/(4\pi|\v x|^{3})$, together with its compensating surface
charge, in which case every subbody is neutral. The two are distinguished
by the surface charges they induce on cuts, which is exactly the information
carried by $\v P$. 
\end{rem}

\begin{rem}[Free charge as a divergence]
\label{rem:Df} Every free charge density is a divergence: there
is a vector field $\v D_{f}$ such that 
\begin{equation}
\nabla\cdot\v D_{f}=r.\label{eq:Df}
\end{equation}
For example, if $r$ is integrable with bounded support (extended
by zero), or more generally a finite measure such as $q\delta_{\v 0}$,
one may take the gradient of its Newtonian potential, 
\begin{equation}
\v D_{f}(\v x)=\frac{1}{4\pi}\int r(\v y)\,\frac{\v x-\v y}{|\v x-\v y|^{3}}\,\dee{V_{\v y}}.\label{eq:Newton}
\end{equation}
The field $\v D_{f}$ is not unique, since any divergence-free field
may be added to it; it is unique among the curl-free fields that vanish
at infinity. By the divergence theorem, 
\begin{equation}
\int_{\gO}r\vph\dV=\int_{\bdry\gO}(\v D_{f}\cdot\v{\nor})\,\vph\dA-\int_{\gO}\v D_{f}\cdot\nabla\vph\dV.\label{eq:freeDf}
\end{equation}

\emph{The point charge.} For $r=q\delta_{\v 0}$, (\ref{eq:Newton})
gives exactly the field (\ref{eq:Dpoint}), and (\ref{eq:freeDf})
is (\ref{eq:Upoint}) rearranged. Thus the same vector field $q\v x/(4\pi|\v x|^{3})$
represents a point charge $q$ at the origin when it is used as $\v D_{f}$,
through $r=\nabla\cdot\v D_{f}$, while, used as $\v P$, it represents
the bound point charge $-q$ together with its compensating surface
charge.

\emph{The two roles are different.} The field $\v P$ is determined
by the charge system, through the surface charges $s_{\gO}=\v P\cdot\v{\nor}$
on all subbodies. The field $\v D_{f}$ is only a potential for $r$,
and it is not determined by the charge system. Equation (\ref{eq:freeDf})
states that the free charge has the energy of the dipole distribution
with density $-\v D_{f}$, whose bound volume charge is $\nabla\cdot\v D_{f}=r$,
together with the additional surface charge $\v D_{f}\cdot\v{\nor}$
on $\bdry\gO$. The latter is the negative of the bound surface charge
$-\v D_{f}\cdot\v{\nor}$ of that dipole distribution, so the two
cancel on every cut. Free charge is therefore the bound volume charge
of the dipole distribution $-\v D_{f}$, stripped of its bound surface
charge. Writing $r=\nabla\cdot\v D_{f}$ does not turn free charge
into dipoles: $r$ is the part of the charge that exposes no charge
on cuts, and its total charge $Q_{\gO}(1)=\int_{\bdry\gO}\v D_{f}\cdot\v{\nor}\dA$,
the flux of $\v D_{f}$ (Gauss's law), need not vanish.
\end{rem}

\section{Mechanical forces on embedded charge distributions}

\label{sec:forces}

\subsection{Embedded charge distributions}

Let $\gO$ be a material body and $\dis:\reals\cross\gO\to\rthree$
a smooth virtual motion with $\dis_{0}=\mathrm{Id}$, configurations
$\dis_{t}$ that are orientation-preserving embeddings, deformation
gradient $F_{t}=\nabla\dis_{t}$, Jacobian $J_{t}=\det F_{t}>0$,
and virtual velocity $\v w=\bdry_{t}\dis_{t}|_{t=0}$. The potential
$\vph$ is held fixed in space.

We say that the charge distribution $(r,\v P)$ is embedded in the
body if the free charge of every material element is conserved and,
as in Section~\ref{sec:dipole}, each dipole is carried as a material
line element. For the free charge $r\dV$ of a volume element this
means $r_{t}\,\dee{V_{t}}=r\dV$, i.e., $r_{t}=r/J_{t}$. For the
dipole moment $\v P\dV$ it means 
\begin{equation}
\v P_{t}\,\dee{V_{t}}=F_{t}\v P\dV,\qquad\text{i.e.,}\qquad\v P_{t}=\frac{1}{J_{t}}F_{t}\v P,
\end{equation}
the Piola transformation. By Nanson's formula, $\v P_{t}\cdot\v{\nor}_{t}\,\dee{A_{t}}=\v P\cdot\v{\nor}\dA$,
so the bound surface charge of every material surface is conserved,
and by the Piola identity so is the bound charge of every material
subbody.

The embedding assumption rules out conduction: the free charge cannot
move relative to the material. So the results below apply to dielectrics
and to free charge fixed in insulating matter. Conductors and currents
would also require a flux of free charge relative to the body, which
we do not consider here.

\subsection{The force functional}

The energy (\ref{eq:genU}) of the convected distribution at time
$t$ is 
\begin{equation}
\begin{split}U_{t} & =\int_{\dis_{t}(\gO)}\big(r_{t}\vph+\v P_{t}\cdot\nabla\vph\big)\,\dee{V_{t}}\\
 & =\int_{\gO}r(\v x)\,\vph(\dis_{t}(\v x))\dV+\int_{\gO}\dis_{t\,i,j}(\v x)\,P_{j}(\v x)\,\vph_{,i}(\dis_{t}(\v x))\dV.
\end{split}
\label{eq:Ut}
\end{equation}
Differentiating at $t=0$ exactly as in (\ref{eq:Fd}), 
\begin{equation}
\frac{d}{dt}U_{t}\Big|_{t=0}=\int_{\gO}\big(r\vph_{,k}w_{k}+w_{i,j}P_{j}\vph_{,i}+P_{i}\vph_{,ik}w_{k}\big)\dV.
\end{equation}
Assuming, as in Section~\ref{sec:dipole}, that the system is conservative,
we define the force functional by $F_{\gO}(\v w):=-\frac{d}{dt}U_{t}|_{t=0}$.
With $\vph_{,i}=-E_{i}$ and $E_{i,k}=E_{k,i}$, 
\begin{equation}
\begin{split}F_{\gO}(\v w) & =\int_{\gO}\big(rE_{k}w_{k}+P_{i}E_{k,i}w_{k}+E_{i}P_{j}w_{i,j}\big)\dV\\
 & =\int_{\gO}\Big(\big(r\v E+(\v P\cdot\nabla)\v E\big)\cdot\v w+(\v E\otimes\v P):\nabla\v w\Big)\dV.
\end{split}
\label{eq:F}
\end{equation}
Comparing the last expression with the results (\ref{eq:Wd2}) and
(\ref{eq:Fd}), the continuum theory may be viewed as the continuum
analog of superposition of embedded point charges and dipoles, both
for the energy and for the force.
\begin{prop}
\label{prop:forcegauge} For every charge system $(r,\v P)$, the
force functional (\ref{eq:F}) is gauge invariant: it is unchanged
when $\vph$ is replaced by $\vph+c$, $c$ a constant. 
\end{prop}

\begin{proof}
By (\ref{eq:Ut}), replacing $\vph$ by $\vph+c$ changes $U_{t}$
by $c\int_{\gO}r\dV$, which does not depend on $t$, since the free
charge is convected. Hence $dU_{t}/dt$ is unchanged. Equivalently,
(\ref{eq:F}) depends on $\vph$ only through $\v E=-\nabla\vph$. 
\end{proof}

So gauge invariance holds at the level of forces for every charge
system, but at the level of energies only for neutral ones. In the
mechanical analogy, the former corresponds to a property that every
theory must have, the latter to the special class of equilibrated
force systems.

\subsection{Interpretation as a force system}

Expression (\ref{eq:F}) is a force system in the sense of Section~\ref{sec:mech},
now for genuine virtual velocities. Its variational stress is 
\begin{equation}
\v r_{m}=r\v E+(\v P\cdot\nabla)\v E,\qquad\sigma_{m}=\v E\otimes\v P,\quad(\sigma_{m})_{ij}=E_{i}P_{j}.
\end{equation}
Since $(\nabla\cdot\sigma_{m})_{i}=(E_{i}P_{j})_{,j}=E_{i,j}P_{j}+E_{i}\nabla\cdot\v P$,
the body force and the surface traction are 
\begin{equation}
\v b_{m}=\v r_{m}-\nabla\cdot\sigma_{m}=(r-\nabla\cdot\v P)\v E=\rho\,\v E,\qquad\v t_{\gO}=\sigma_{m}\v{\nor}=(\v P\cdot\v{\nor})\v E=s_{\gO}\v E,\label{eq:bt}
\end{equation}
and 
\begin{equation}
F_{\gO}(\v w)=\int_{\gO}\rho\,\v E\cdot\v w\dV+\int_{\bdry\gO}s_{\gO}\,\v E\cdot\v w\dA,\qquad\rho=r+\rho_{b}.\label{eq:Fbt}
\end{equation}
The electric field exerts the Coulomb force on the total volume charge,
free and bound, and on the bound surface charge exposed on $\bdry\gO$.
The two representations, (\ref{eq:F}) by a variational stress and
(\ref{eq:Fbt}) by a body force and a traction, are equivalent. They
are the continuum counterparts of the two computations of Section~\ref{sec:dipole}:
the first through the dipole moment, the second through the individual
charges.

\textbf{Contrast between the two levels.} At the level of potentials,
gauge invariance of the energy characterizes $r=0$. At the mechanical
level, the zeroth-order term $\v r_{m}=r\v E+(\v P\cdot\nabla)\v E$
is in general nonzero even for a neutral system, and $F_{\gO}(\v c)=\v c\cdot\int_{\gO}\v r_{m}\dV\neq0$.
This is as it should be: the electric field does exert a net force
on matter, and this force must be balanced by other forces, e.g.,
by the mechanical stress in the material or by external supports.
By Proposition~\ref{prop:mechbalance}, the loading by the electric
field cannot be represented by a stress acting on $\nabla\v w$ alone.

\textbf{The active stress and the couple density.} The tensor $\sigma_{m}=\v E\otimes\v P$
acts on the velocity gradient, so it is a stress. It is an active
stress exerted by the electric field: it is neither the stress in
the material nor the Maxwell stress, and it is not symmetric. For
a rigid rotation $\v w=\v{\go}\cross\v x$, $w_{i,j}=\eps_{ikj}\omega_{k}$
and 
\begin{equation}
E_{i}P_{j}w_{i,j}=\v{\go}\cdot(\v P\cross\v E),
\end{equation}
so the skew-symmetric part of $\sigma_{m}$ is the couple density
$\v P\cross\v E$, the continuum counterpart of the couple $\v d\cross\v E$
of Section~\ref{sec:dipole}.

\section{Examples}

\label{sec:examples}
\begin{example}[Total force and torque]
\label{ex:total} For a uniform virtual velocity, (\ref{eq:F}) and
(\ref{eq:Fbt}) give the total force 
\begin{equation}
\v F=\int_{\gO}\big(r\v E+(\v P\cdot\nabla)\v E\big)\dV=\int_{\gO}\rho\,\v E\dV+\int_{\bdry\gO}s_{\gO}\v E\dA,
\end{equation}
the standard force on a charged, polarized body. For a rigid rotation
about the origin, the total torque is 
\begin{equation}
\v M=\int_{\gO}\big[\v x\cross\big(r\v E+(\v P\cdot\nabla)\v E\big)+\v P\cross\v E\big]\dV.
\end{equation}
\end{example}

\begin{example}[Uniform field]
If $\v E=\v E_{0}$ is uniform, then $(\v P\cdot\nabla)\v E=\v 0$,
and 
\[
\v F=Q\,\v E_{0},\qquad\v M=\Big(\int_{\gO}r\,\v x\dV+\v p\Big)\cross\v E_{0},\qquad Q=\int_{\gO}r\dV,\quad\v p=\int_{\gO}\v P\dV.
\]
The force acts on the net (free) charge only. The torque involves
the total dipole moment of the body, $\int_{\gO}r\v x\dV+\v p$, the
first term from the free charge and the second from the dipoles. For
a neutral body, $\v F=\v 0$ and $\v M=\v p\cross\v E_{0}$, the continuum
counterparts of the vanishing force and the couple $\v d\cross\v E$
on a single dipole. 
\end{example}

\begin{example}[An isolated point charge]
Let $r=q\delta_{\v 0}$ and $\v P=\v 0$, with the origin in the
interior of $\gO$. Then $Q_{\gO}(\vph)=q\vph(\v 0)$ and (\ref{eq:F})
gives $F_{\gO}(\v w)=q\v E(\v 0)\cdot\v w(\v 0)$, the Coulomb force.
The energy is not gauge invariant; the force is. By Remark~\ref{rem:Df},
$r=\nabla\cdot\v D_{f}$ with $\v D_{f}$ given by (\ref{eq:Dpoint}). 
\end{example}

\begin{example}[Uniformly polarized body in a field with uniform gradient]
Let $r=0$, $\v P=\v P_{0}$ be uniform in $\gO$, and $\v E(\v x)=\v E_{0}+G\v x$
with $G$ a constant symmetric matrix (so that $\nabla\cross\v E=\v 0$).
Then $\rho_{b}=0$, the whole bound charge is on the surface, $s_{\gO}=\v P_{0}\cdot\v{\nor}$,
and 
\begin{equation}
\v F=|\gO|\,G\v P_{0},
\end{equation}
the force $(\v d\cdot\nabla)\v E$ on the total dipole $\v d=|\gO|\v P_{0}$. 
\end{example}

\begin{example}[A smoothly tapered dipole distribution]
\label{ex:bump} Let $\vph=-\tfrac{1}{2}kx^{2}_{1}$, so that $\v E=(kx_{1},0,0)$,
and let $r=0$ and $\v P=f(\v x)\v e_{1}$, where $f\geq0$ is smooth
and compactly supported in the interior of $\gO$. Then $\v P\cdot\v{\nor}=0$
on $\bdry\gO$, and by (\ref{eq:F}) 
\begin{equation}
F_{1}=\int_{\gO}P_{1}E_{1,1}\dV=k\int_{\gO}f\dV,
\end{equation}
the attraction of the dipoles toward the region of stronger electric
field for $k>0$. There is no surface charge on $\bdry\gO$, and in
the representation (\ref{eq:Fbt}) the whole force acts on the bound
volume charge $\rho_{b}=-\bdry_{1}f$. 
\end{example}

\begin{example}[Forces on the hedgehog polarization]
\label{ex:point} Let $r=0$ and $\v P$ be given by (\ref{eq:Dpoint}),
embedded in a body $\gO$ containing the origin with $\v 0\notin\bdry\gO$,
and let $\v E$ be a smooth external field. Since $|\v P|=O(|\v x|^{-2})$
and $\nabla\v E$ is bounded near the origin, the force functional
(\ref{eq:F}) is well defined. The excision argument of Proposition~\ref{prop:point},
applied to $\nabla\cdot\big((\v E\cdot\v w)\v P\big)=\v P\cdot\nabla(\v E\cdot\v w)$
on $\gO\setminus B_{\eps}$, gives 
\begin{equation}
F_{\gO}(\v w)=\int_{\gO}\v P\cdot\nabla(\v E\cdot\v w)\dV=-q\,\v E(\v 0)\cdot\v w(\v 0)+\int_{\bdry\gO}(\v P\cdot\v{\nor})\,\v E\cdot\v w\dA,
\end{equation}
the singular counterpart of (\ref{eq:Fbt}). For a uniform velocity,
the total force 
\begin{equation}
\v F=-q\,\v E(\v 0)+\int_{\bdry\gO}(\v P\cdot\v{\nor})\,\v E\dA
\end{equation}
is the force on the bound point charge $-q$ plus the force on the
compensating surface charge. It vanishes when $\v E$ is uniform,
as for any neutral body. 
\end{example}

\section{Concluding remarks}

\label{sec:remarks}

\textbf{Summary.} Electrostatics is the stress theory we get by replacing
velocities with scalar potentials. Cauchy's theorem represents every
charge system by a pair $(r,\v P)$, the analog of a variational stress,
which splits the charge uniquely into free charge $r$ and a dipole
distribution $\v P$ carrying the bound charges $-\nabla\cdot\v P$
and $\v P\cdot\v{\nor}$. Gauge invariance of the energy is not a
postulate. It plays the role of the balance of forces on every subbody:
it holds if and only if every subbody is neutral, if and only if $r=0$,
and if and only if the energy is $-\int\v P\cdot\v E\dV$. A point
charge may be free, or bound together with its compensating charge.
The mechanical forces on embedded charges are gauge invariant for
every charge system. They consist of the Coulomb force on the total
charge, free and bound, or, equivalently, of the force $r\v E$ on
the free charge, the dipole force density $(\v P\cdot\nabla)\v E$,
and the asymmetric active stress $\v E\otimes\v P$, whose skew-symmetric
part is the couple density $\v P\cross\v E$.

The equations of electrostatics in a polarizable medium are (e.g.,
{[}Jac99, Zan13{]}) 
\begin{equation}
\nabla\cross\v E=\v 0,\qquad\nabla\cdot\v D=\rho_{f},\qquad\v D=\eps_{0}\v E+\v P,\label{eq:maxwell}
\end{equation}
where $\v D$ is the electric displacement, $\rho_{f}$ the free charge
density and $\v P$ the polarization. We have not used any of these
equations. We have treated the potential $\vph$ and the charge system
$(r,\v P)$ as independent: $\vph$ is a potential, or a variation
of it, applied to a fixed charge system, just as, in stress theory,
a velocity field is applied to a fixed force system regardless of
the constitutive properties of the body. The framework is thus a theory
of the interaction between a given charge system and a given potential.
Maxwell's equations come in once we require the potential to be the
one generated by the charges. They then play the role of the remaining
ingredients of a boundary value problem of continuum mechanics.

Since $\v E=-\nabla\vph$, the first equation of (\ref{eq:maxwell})
holds identically. It is the analog of the compatibility of a strain
field derived from a displacement field: here the potential is the
primary field, and $\nabla\vph$ plays the role of the velocity gradient.

\textbf{Generalization.} The analogy extends to magnetostatics, where
the potential is a vector field $\v A$ and the gauge transformations
$\v A\mapsto\v A+\nabla\gl$ play the role of rigid motions. There,
the assumption that a gauge transformation changes the energy only
through the values of $\nabla\gl$, and not through its second derivatives,
is equivalent to the skew-symmetry of the ``stress'' acting on $\nabla\v A$,
which is then represented by a vector field $\v M$. Gauge invariance
of the energy then characterizes the absence of free currents, in
which case the energy is $-\int\v M\cdot\v B\dV$. More generally,
for $p$-form potentials on a manifold, the variational stress has,
under the analogous assumption, the form $\mathfrak{J}_{f}\wedge\alpha+(-1)^{n-p-1}\mathfrak{g}\wedge d\alpha$,
with free source $\mathfrak{J}_{f}$ and an $(n-p-1)$-form $\mathfrak{g}$,
and gauge invariance of the energy characterizes the case $\mathfrak{J}_{f}=0$.
These cases are treated in the companion papers {[}GMS26a{]} and {[}GMS26b{]}.

\section*{Acknowledgment}

The authors are grateful to Claude (Anthropic), an AI system, which
contributed substantially to the development of the arguments, to
the derivations, and to the preparation of this manuscript, and to
the many colleagues with whom earlier versions of the manuscript were
discussed.

\end{document}